 \documentclass[12pt,draftcls,onecolumn,twoside]{IEEEtran}
\usepackage{acronym,mathrsfs,dsfont,amssymb}  
\usepackage{color,cite}  
\usepackage{graphicx}%
\usepackage{physics}
\usepackage{xtab,booktabs}
\usepackage[tableposition=top]{caption}
\usepackage{longtable}
\usepackage{colortbl,pgfplotstable}
\usepackage{caption}
\usepackage{subcaption}

\newtheorem{theorem}{Theorem}
\newtheorem{lemma}{Lemma}

\newcommand{\Set}[1]{\mathcal{#1}}

\acrodef{epr}[EPR]{Einstein-Podolsky-Rosen}
\newcommand\p{\lambda}

\title{Entanglement Swapping for Repeater Chains with Finite Memory Sizes}
\author{
	\vspace{0.2cm}
    Wenhan~Dai
    and
    Don~Towsley  
\thanks{W.~Dai is with the College of Information and Computer Sciences, University of Massachusetts, Amherst and the Quantum Photonics Laboratory, Massachusetts Institute of Technology (e-mail:  whdai@cs.umass.edu).}
	\thanks{D.~Towsley is with the College of Information and Computer Sciences, University of Massachusetts, Amherst (e-mail:  towsley@cs.umass.edu).} 
    }

\begin{document}

\maketitle

\begin{abstract}
We develop entanglement swapping protocols and memory allocation methods for quantum repeater chains. Unlike most of the existing studies, the memory size of each quantum repeater in this work is a parameter that can be optimized. Based on Markov chain modeling of the entanglement distribution process, we determine the trade-off between the entanglement distribution rate and the memory size. We then propose three memory allocation methods that achieve entanglement distribution rates decaying polynomially with respect to the distance while using a constant average number of memories per node. We also quantify the  average  number of memories required due to classical communication delay, as well as the delay of entanglement distribution. Our results show that a moderate memory size suffices to achieve a polynomial decay of entanglement distribution rate with respect to the distance, which is the scaling achieved by the optimal protocol even with infinite memories at each node. 
\end{abstract}

\acresetall		

\section{Introduction}

The quantum Internet can revolutionize many applications in the field of communication, computation, and sensing \cite{WehElkHan:18}. One important service of the quantum Internet is its capability to distribute entanglement among end users. However, entanglement distribution rate  decays exponentially with respect to the distance over optical fiber \cite{PirLauOttBan:17}. To circumvent the exponential decay of rate with respect to distance, quantum repeaters need to be inserted between end users. By generating entanglement between quantum repeaters with quantum channels and entanglement swapping at the quantum repeaters, it is possible to achieve entanglement distribution rate that decays much slower than exponentially obtained by direct transmission. 



However, the quantum repeaters may not be perfect, e.g., entanglement swapping may have errors or succeed probabilistically. If entanglement swapping has errors, one can apply entanglement distillation or error correction to improve the fidelity of entanglement \cite{BriDurCirZol:98,JiaTayKhaLuk:07, JiaTayNemMunMetLuk:09,ZwePirDunBriDur:18}. Together with a nesting method, an entanglement distribution rate that decays polynomially with respect to the distance can be achieved \cite{BriDurCirZol:98,JiaTayKhaLuk:07, JiaTayNemMunMetLuk:09}. If entanglement swapping fails with a nonzero probability, the entanglement distribution rate may vary significantly depending on the setting, such as the number of memories and the lifetime of qubits \cite{DaiPenWin:J20b,PanKroTowTasJiaBasEngGuh:19,ShcSchLoo:19,ShcLoo:21,JiaTayLuk:07b,CooBraElk:21,BraCooElk:20, GooElkWeh:21}. For one extreme case where each node has infinite memories and qubits have sufficiently long lifetime, the optimal entanglement distribution rate can be obtained in a closed form for homogeneous repeater chains \cite{DaiPenWin:J20b}. For the other extreme case where each node has one memory per link and qubits have lifetime of one time slot, protocols for quantum repeaters have been developed  by exploiting the diversity of multiple paths in the network \cite{PanKroTowTasJiaBasEngGuh:19}. Although studies in these two extreme cases offer important insight into the design of quantum networks in the NISQ era \cite{Pre:18}, it is much more common to encounter intermediate cases where nodes have a finite number of memories and reasonably long lifetime. Existing studies generally consider that each node has one memory per link \cite{ShcSchLoo:19, ShcLoo:21,JiaTayLuk:07b,CooBraElk:21}, and it is unclear how increased memory size can improve the entanglement generation rate. Moreover, in most existing entanglement swapping protocols, a node's operation may depend on the results of entanglement swapping at another node that is far away. The classical communication delay cannot be ignored and it may be detrimental to the entanglement distribution rate. However, the effects of classical communication delay in the entanglement distribution rate and memory size requirement are not well studied. In \cite{ShcSchLoo:19, ShcLoo:21}, the classical communication delay is considered for determining the expected waiting time in quantum repeaters, but explicit expressions are provided for only a few links. For a general case, the computation complexity grows fast with respect to the number of links.



To the best of the authors’ knowledge, it remains unknown how the entanglement distribution rate depends on the memory size and classical communication delay. A fundamental question related to entanglement distribution is, for a targeted entanglement distribution rate, how many memories are needed at each node in a network when classical communication delay is considered? 
The answer to this question will guide the design of quantum repeaters, the deployment of quantum nodes, and the implementation of entanglement swapping protocols.

The goal of this manuscript is to determine the effects of memory size and classical communication delay on entanglement distribution rates, and develop memory allocation methods based on these effects. Key contributions of this work are as follows.
\begin{itemize}
    \item We develop a modified doubling method that schedules entanglement swapping operations and determine an approximation of the entanglement distribution rate as a function of the number of memories.
    \item We proposed three memory allocation methods. These methods achieve an entanglement distribution rate decaying polynomially with respect to the distance while using a constant average number of memories per node.
    \item We determine the expected number of memories per node required due to classical communication delay, and show that this number does not rely on the distance between end users.
    \item We quantify the delay of entanglement distribution, i.e., the  maximum  time  that a qubit  stays  in  the  memory during  the  entanglement  distribution  process.
\end{itemize}


\section{System Model}
\begin{figure}[t]
\center	
\includegraphics[width=0.65\linewidth, draft=false]{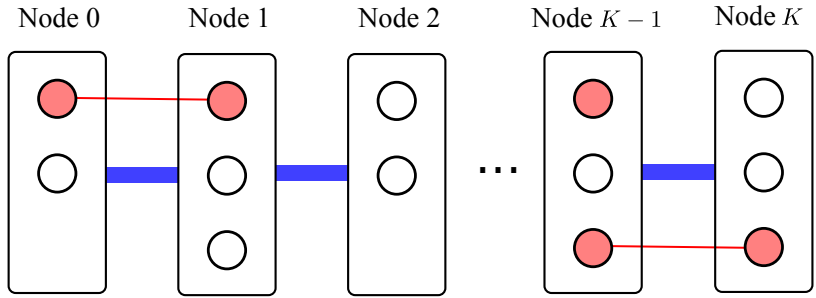}
\caption{Illustration of repeater chains. Circles represent memories for storing entangled qubits. Red solid circle represent entangled qubits. Blue lines represent quantum channels.}
\label{fig:illustration_repeater_chain}
\end{figure}

Consider a repeater chain consisting of $K$ nodes and quantum channels illustrated in Fig.~\ref{fig:illustration_repeater_chain}. We assume $K$ is power of 2 in this paper, i.e., $K=2^k$, $k\in \mathbb{N}^*$, but the results are applied to the case with a general $K$. The nodes are labelled as 0, 1, $\dotsc$, $K$. There exist a quantum channel that connects node $i$ and $i+1$, $i=0,1,2,\dotsc, K-1$. The length of these channels are the same. A quantum node may consist multiple memories to store entangled qubits. The goal is to maximize the distribution rate of \ac{epr} pair between two end node, nodes 0 and $K$, via entanglement generation and entanglement swapping.


We consider a continuous time model, where the successful entanglement generation between neighboring nodes follows a Poisson process with rate $\p$. Consequently, two entanglement generation attempts cannot succeed simultaneously. Any node $i=1,2,\dotsc, K-1$ may perform entanglement swapping that consumes two \ac{epr} pairs $\ket{\Psi_{ij}}$ and $\ket{\Psi_{ij'}}$ to distribute an \ac{epr} pair $\ket{\Psi_{jj'}}$. Such entanglement operation succeeds with probability $q$. If it succeeds, we assume that entanglement swapping has no errors; if it fails, we assume no entanglement is obtained.

Note that entanglement swapping at certain nodes may depend on the operation at other nodes. This incurs the delay due to the delay to transmit classical messages and consequently extra memory requirement. We will consider no classical communication delay for now and leave the discussion of classical communication in Section \ref{sec:ccd_buffersize}.

\section{Temporal Multiplexing}
In temporal multiplexing, the generated entangled pairs can be stored in the memory  so that these pairs can be used in the future \cite{DhaPatKroGuh:21}. The performance of temporal multiplexing depends on entanglement swapping protocols. It is shown that the ``doubling method'' can achieve the optimal entanglement generation rate with infinite memories and qubit lifetime \cite{DaiPenWin:J20b}. 

\subsection{Level $i$ Entanglement and Doubling Method}
\begin{figure}[t]
\center	
\includegraphics[width=0.65\linewidth, draft=false]{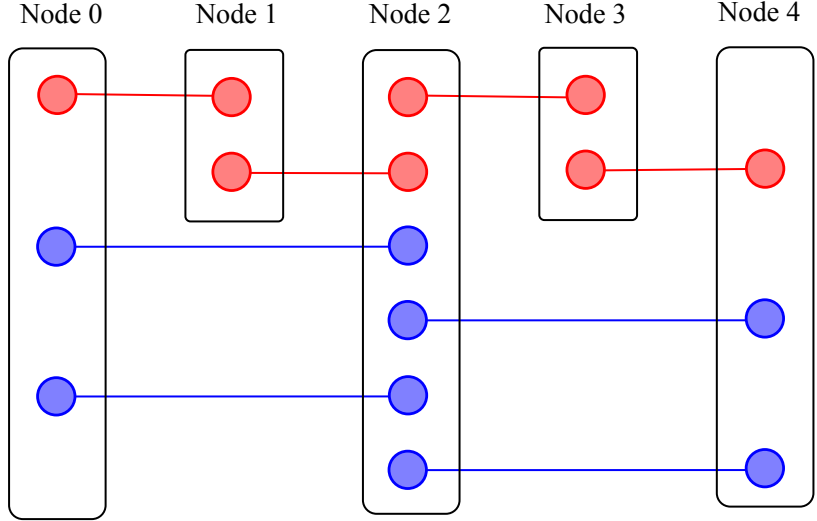}
\caption{Illustration of level $k$ entanglement. The red lines and red circles represent entanglement that spans one channel, and such entanglement is referred to as level 0 entanglement. Similarly, blue lines and blue circles represent level 1 entanglement. In this figure, $B_0=1$ and $B_1=2$.}
\label{fig:illustration_temporal_modified_doubling}
\end{figure}

To explain the doubling method, we need to introduce the concept of level $i$ entanglement. As shown in Fig. \ref{fig:illustration_temporal_modified_doubling}, we refer to entangled pair that spans $2^i$ channels as level $i$ entanglement, where $i=0,1,2,\dotsc,k$. The doubling method only allows entanglement swapping at node $n\in\{2^i, 2^i *3, \dotsc, K-2^i\}$, $i=1,2,\dots, k-1$: whenever there are entanglement $\ket{\Psi_{(n-2^i) n}}$ and $\ket{\Psi_{n(n+2^i)}}$, node $n$ swaps entanglement to create a level $i+1$ entanglement $\ket{\Psi_{(n-2^i) (n+2^i)}}$. If each node has infinite memories, then the optimal entanglement distribution rate $pq^k$ can be achieved. On the other hand, even if there is only one memory per link at each node, the doubling method is shown to achieve the entanglement generation rate of $p(2q/3)^k$ \cite{CooBraElk:21}. In this section, we study an intermediate case, where there are finite memories at each node, and aim to develop memory allocation methods that can maximize the entanglement distribution rate.


As shown in  Fig. \ref{fig:illustration_temporal_modified_doubling}, we assign different number of memories to store different levels of entanglement. In particular, nodes 1, 2, $\dotsc$, $K-1$ are allocated with $2B_0$ memories to store level 0 entanglement, and at node $n\in\{1,2,\dotsc,K-1\}$, $B_0$ memories store entanglement $\ket{\Psi_{(n-1) n}}$ and the other $B_0$ memories store entanglement $\ket{\Psi_{n (n+1)}}$. Similarly, nodes $2^i$, $2^i *2$, $\dotsc$, $K-2^i$ are allocated with $2B_i$ memories to store level $i$ entanglement, $i=1,2,\dotsc, k-1$, and at node $n\in\{2^i$, $2^i *2, \dotsc, K-2^i\}$, $B_i$ memories store entanglement $\ket{\Psi_{(n-2^i) n}}$ and the other $B_i$ memories store entanglement $\ket{\Psi_{n (n+2^i)}}$. 

We consider a modified doubling method that works as follows. Each channel keeps generating level 0 entanglement until the $B_0$ memories are full. For a node $n\in\{2^i, 2^i *3, \dotsc, K-2^i\}$, $i=1,2,\dots, k-1$, whenever there are entanglement $\ket{\Psi_{(n-2^i) n}}$ and $\ket{\Psi_{n(n+2^i)}}$, node $n$ swaps entanglement to create a level $i+1$ entanglement $\ket{\Psi_{(n-2^i) (n+2^i)}}$ until the $B_{i+1}$ memories are full. When the memories of some entanglement are full, then the protocol can choose to throw certain lower level entanglement or stops  generating lower level entanglement. The following theorem shows the trade-off between memory size and entanglement distribution rate achieved by the modified doubling entanglement swapping protocol for small $q$.

\begin{theorem}\label{thm:rate_approximation}
Let $R_k(\p, q)$ denote the entanglement distribution rate achieved by a modified doubling entanglement swapping protocol. Then
\begin{align*}
    \lim_{q\rightarrow 0} \frac{\underline{R}_k(p,q)}{ R_k(p,q) } = 1.
\end{align*}
where 
\begin{align}\label{eq:R_small_q}
\underline{R}_k(p,q) =    \p q^k \prod_{i=0}^{k-1}\frac{2B_i}{2B_i+1}.
\end{align}
\end{theorem}
\begin{IEEEproof}
See Appendix \ref{apd:proof_rate_approximation}.
\end{IEEEproof}


Theorem \ref{thm:rate_approximation} shows that for sufficiently small $q$, the entanglement distribution rate achieved by a modified doubling entanglement swapping protocol can be accurately estimated by $\underline{R}_k(p,q)$. In fact, analysis for small scale networks (e.g., $K=4$) and numerical results show that  $\underline{R}_k(p,q)$ may be a lower bound for $R_k(\p, q)$. In the rest of this section, we will use \eqref{eq:R_small_q} as an approximation of the entanglement distribution rate for any $q\in(0, 1]$ and analyze the memory size with respect to the distance (or equivalently, $K$).

\subsection{Memory Allocation Methods}
We now consider two memory allocation methods. The first method is the constant memory allocation method. If we set $B_i$ to be constant, i.e., $B_i =B>0$, $i=0,1,2,\dotsc,k-1$, then the entanglement distribution rate becomes $\p(2Bq/(2B+1))^k$. Note that such a rate decays polynomially with respect to the distance. The average number of memories is upper bounded by
a constant $4B$ because
\begin{align*}
    2B + 2B /2 + 2B/2^2 + ... + 2B/2^k \leq 4B
\end{align*}
and the maximum number of memory among the $K+1$ nodes is $2kB$, which grows logarithmally with respect to the distance. 


We can use another memory allocation method, referred to as the exponential memory allocation, to achieve a near optimal entanglement distribution rate. Let $\delta>0$ denote a preset constant. The goal is to achieve an entanglement distribution rate $\p q^k (1-\delta)$ using the fewest memories. In fact, we can set $B_i = \gamma^{i+i_0}$, $i=0, 1,2,\dots, k-1 $, where $\gamma\in(1,2)$ and $i_0\in\mathbb{N}^*$ are constants chosen later. The average number of memories is
\begin{align*}
    \frac{1}{K}\sum_{i=0}^{k-1} 2B_i \cdot (2^{k-i}-1)& \leq   \frac{1}{2^k}\sum_{i=0}^{k-1} 2\gamma^{i+i_0}\cdot 2^{k-i}\\
    &\leq 2\gamma^{i_0}\sum_{i=0}^{k-1}(\gamma/2)^i\leq \frac{2\gamma^{i_0}}{1-\gamma/2}
\end{align*}
which is a constant independent of the distance. Regarding the entanglement distribution rate, Theorem \ref{thm:rate_approximation} implies that the rate can be lower bounded by
\begin{align*}
    pq^k \prod_{i=0}^{k-1}\frac{2\gamma^{i+i_0}}{2\gamma^{i+i_0}+1}.
\end{align*}

\begin{lemma}\label{lemma:i_0_selection}
There exists $i_0\in\mathbb{N}^*$ such that for any $k\in \mathbb{N}^*$
\begin{align*}
    \prod_{i=0}^{k-1}\frac{2\gamma^{i+i_0}}{2\gamma^{i+i_0}+1} \geq 1-\delta.
\end{align*}
\end{lemma}
\begin{IEEEproof}
Consider the following sequence:
\begin{align*}
    a_i = \log{\frac{2\gamma^{i}}{2\gamma^{i}+1}}.
\end{align*}
Evidently, $a_i<0$. Since 
\begin{align*}
    \lim_{i\xrightarrow{}\infty}\frac{a_{i+1}}{a_{i}} = \lim_{i\xrightarrow{}\infty}\frac{ 1/(2\gamma^{i+1}+1)}{ 1/(2\gamma^{i}+1) } =1/\gamma < 1,
\end{align*}
we have
\begin{align*}
    \sum_{i=0}^{\infty}a_i >-\infty.
\end{align*}
Therefore, there exists $i_0\in \mathbb{N}^*$ such that
\begin{align}\label{eq:finite_sum_a}
    \sum_{i=i_0}^\infty a_i \geq \log(1-\delta).
\end{align}
We then have the desired result because
\begin{align*}
    \prod_{i=0}^{k-1}\frac{2\gamma^{i+i_0}}{2\gamma^{i+i_0}+1} \geq \prod_{i=0}^{\infty}\frac{2\gamma^{i+i_0}}{2\gamma^{i+i_0}+1}\geq \exp\big\{\sum_{i=i_0}^\infty a_i\big\} \geq 1-\delta
\end{align*}
where the last inequality is due to \eqref{eq:finite_sum_a}. 
\end{IEEEproof}

The results above shows that the temporal multiplexing achieves the entanglement distribution rate $\p q^k(1-\delta)$ and the average number of memories is a constant independent of the distance. The parameter $\gamma$ can be an arbitrary number in $(1,2)$, and $i_0$ depends on $\delta$ and $\gamma$ suggested in Lemma \ref{lemma:i_0_selection}. Note that the maximum memory size is
\begin{align*}
    2\sum_{i=0}^{k-1} B_i = 2\sum_{i=0}^{k-1} \gamma^{i+i_0} = 2\gamma^{i_0}\frac{\gamma^k-1}{\gamma-1}\leq \frac{2\gamma^{i_0}}{\gamma-1}\gamma^k = \frac{2\gamma^{i_0}}{\gamma-1} K^{\log{\gamma}}
\end{align*}
Since $\gamma<2$, the value above grows sub linearly with respect to $K$.

\subsection{More Efficient Use of Memory: Cognitive Method}
The memory allocation methods require that $B_i$ memories are reserved to store level $i$ entanglement, and this may not be the most efficient way of using memories. For example, if the $B_0$ memories that store level 0 entanglement are full but there are unused memories that store higher level entanglement, the memory allocation method in previous sections do not use these free memories. We next propose a cognitive method that allows using these memories. Suppose each node is allocated with $2B$ memories, and for each link, $B$ memories are allocated. The doubling entanglement swapping protocol is used. Higher level entanglement has priority to use memories. Whenever there are free memories, lower level entanglement can be generated through channels or entanglement swapping. 

This method is referred to the cognitive method because the use of memory is similar to spectrum use in cognitive radio, where primary users have priority to use spectrum and if spectrum is not occupied by primary users, the secondary users can use the spectrum.


\begin{theorem}\label{thm:rate_approximation_cons}
Let $E(p,q,B,i)$ denote the entanglement distribution rate achieved by the cognitive memory allocation method for a repeater with $2^i$ channels and $2B$ memories ($B$ memories per link). Then
\begin{align*}
    \lim_{q\rightarrow 0} \frac{E(p,q,B,i)}{\underline{E}(p,q,B,i)} \geq 1
\end{align*}
where $\underline{E}(p,q,B,i)=pq^i f_{B,i}$, in which $f_{B,i}$ is obtained recursively as follows. For $i=0$, $f_{B,0} = 2B/(2B+1)$. For $i>0$,
\begin{align*}
   \frac{f_{B, i}}{f_{B, i-1}} = 1-\bigg[ 1+2\sum_{j =0} ^{B-1}\frac{\prod_{j_0=0}^j f_{B-j_0, i-1}}{(f_{B,i-1})^{j+1}}\bigg]^{-1}.
\end{align*}

\end{theorem}
\begin{IEEEproof}
See Appendix \ref{apd:proof_rate_approximation_cons}.
\end{IEEEproof}

As a special case where $B=1$, one can see that
\begin{align*}
   \frac{f_{B, i}}{f_{B, i-1}} = 1-\bigg[ 1+2 \frac{  f_{1, i-1}}{ f_{1,i-1} }\bigg]^{-1}=\frac{2}{3}.
\end{align*}
Consequently, $f_{1,i} = (2/3)^i$ and $\underline{E}(p,q,1,i)=p(2q/3)^i$. This is consistent with Proposition 3(d) in \cite{CooBraElk:21}.


\section{Classical Communication Delay}\label{sec:ccd_buffersize}
The protocols above rely on classical communication that informs certain nodes whether entanglement swapping is successful. For example, consider a simple case where $K=2$. Suppose node 1 swaps entanglement to create $\ket{\Psi_{02}}$ by  consuming $\ket{\Psi_{01}}$ and $\ket{\Psi_{12}}$. This operation can be success or failure. Before this result is received by node 0, node 0 has to hold its part of $\ket{\Psi_{01}}$. This increases the memory size requirement. To emphasize the different roles of memory, we refer to the $2B_i$, $i=0,1,2,\dotsc, k-1$ memories in previous section as \emph{queuing delay memory}, and the memories that store the entanglement during the time waiting for the entanglement swapping result as \emph{communication delay memory}.

We now determine the size of communication delay memory. Without loss of generality, consider that node $2^{n-1}$ swaps entanglement to create $\ket{\Psi_{02^{n}}}$. It takes time $2^{n-1}d/c$ to send the entanglement swapping result to node 0 and $2^n$, where $c$ is the speed of classical communication. Nodes 0 and $2^n$ need to keep their part of $\ket{\Psi_{02^{n-1}}}$ and $\ket{\Psi_{2^{n-1}2^n}}$ for $2^{n-1}d/c$ due to classical communication delay. Suppose the entanglement swapping protocol achieves the rate in Theorem \ref{thm:rate_approximation}. Little's law shows that the expected number of qubits stored in the communication delay memory of node 0 and node $2^{n}$ is upper bounded by
\begin{align*}
  \Big(  \p q^{n-1}\prod_{i=0}^{n-2} \frac{2B_i}{2B_i+1} \Big)\frac{2^{n-1}d}{c}.
\end{align*}
Then the expected number of qubits stored in the communication delay memory average over all the nodes is upper bounded by
\begin{align*}
   \frac{1}{2^k} \sum_{n=1}^k2^{1+
   k-n}\Big(  \p q^{n-1}\prod_{i=0}^{n-2} \frac{2B_i}{2B_i+1} \Big)\frac{2^{n-1}d}{c} &\leq \frac{1}{2^k} \sum_{n=1}^k2^{1+k-n} \p q^{n-1} \frac{2^{n-1}d}{c}\\
   & = \sum_{n=1}^k \frac{\p q^{n-1}d}{c} =\frac{\p d(1-q^k)}{c(1-q)}\leq \frac{\p d}{c(1-q)}
\end{align*}
which is a constant that does not depend on the distance. Therefore, the entanglement distribution rate does not change significantly by adding communication delay memories, the number of which remains constant per node. Moreover, the expected number of qubits stored in the communication delay memory in any node is upper bounded by
\begin{align*}
    \sum_{n=1}^k\Big(\p q^{n-1}\prod_{i=0}^{n-2}\frac{2B_i}{2B_i+1} \Big)\frac{2^{n-1}d}{c} \leq \sum_{n=1}^k \p q^{n-1}\frac{2^{n-1}d}{c}=\frac{\p d}{c}\frac{1-(2q)^k}{1-2q}.
\end{align*}
One can see that if $q<1/2$, this quantity is upper bounded by a constant $\p d/(c(1-q))$; if $q=1/2$, this quantity is $\p dk/c$, which grows logarithmally with respect to the distance as queuing delay memory with $B_i$ being constant. If $q>1/2$, one can discard some entanglement so that communication delay memory size is constant per node while still achieving entanglement distribution rate that decays polynomially with respect to the distance.

\section{Delay of Entanglement Distribution}
The fidelity of the final level $k$ entanglement depends on the delay of entanglement distribution.
In this manuscript, we refer to delay as the maximum time that a qubit stays in the memory during the entanglement distribution process. As mentioned in Section \ref{sec:ccd_buffersize}, there are two types of delay, namely, queuing delay and communication delay. 



Without loss of generality, we consider the node $2^n$, $n\in\{0,1,2,\dotsc, k\}$ and the time that level $i$ entanglement ($i=0,1,2,\dotsc,n$) stays in the memory of node $2^n$. The classical communication delay for level $i$ entanglement is $2^{i-1}d/c$ because of waiting for entanglement swapping results from node $2^n-2^i$. The average queuing delay can be obtained from Little's law as follows.
\begin{align*}
    T_{\mathrm{QD}}(i) = \frac{B_i/2}{\p q^{i-1}\prod_{m=0}^{i-2}2B_m/(2B_m+1)}
\end{align*}
where $T_{\mathrm{QD}}(i)$ denotes the average queuing delay of level $i$ entanglement at node $2^n$, $B_i/2$ is the average queue length of level $i$ entanglement, and $\p q^{i-1}\prod_{m=0}^{i-2}2B_m/(2B_m+1)$ denotes the arrival rate of level $i$ entanglement.

Suppose we use the constant memory allocation, and the average queuing delay becomes
\begin{align*}
    \frac{B}{2\p}\bigg(\frac{2B+1}{2Bq}\bigg)^{i-1}.
\end{align*}
The average delay of level $k$ entanglement is
\begin{align*}
    \frac{B}{2\p}\bigg(\frac{2B+1}{2Bq}\bigg)^{k-1}+2^{k-1}d/c.
\end{align*}
If $(2B+1)/(2Bq)<2$, the delay is $O(K)$.


Suppose we use the exponential memory allocation, and the average queuing delay becomes
\begin{align*}
    \frac{\gamma^{i+i_0}}{2\p q^{i-1}} \prod_{m=0}^{i-2}\frac{2B_m+1}{2B_m} \leq \frac{\gamma^{i+i_0}}{2(1-\delta)\p q^{i-1}}  
\end{align*}
and the average delay of level $k$ entanglement is
\begin{align*}
    \frac{q\gamma^{i_0}}{2\p(1-\delta)}\bigg(\frac{\gamma}{q}\bigg)^{k-1}+2^{k-1}d/c.
\end{align*}
If $\gamma/q<2$, the delay is $O(K)$.






\section{Conclusion}
In this work, we investigate entanglement distribution for quantum repeater chains with finite memory sizes. We develop a modified doubling method that schedules entanglement swapping operations and determine an approximation of the entanglement distribution rate as a function of the number of memories. Based on such approximation, we propose a constant memory allocation method and an exponential memory allocation method. Both methods use constant average memories per node and achieve entanglement distribution rates decaying polynomially with respect to the distance. We also propose a cognitive memory allocation method that can use the memory more efficiently. 

We determine the expected  number of  memories per node required due to classical communication delay. This number does not rely on the distance between end users. Moreover, we quantify the delay of entanglement distribution, i.e., the  maximum  time  that a qubit  stays  in  the  memory during  the  entanglement  distribution  process. If $q>1/2$, then the average delay of entanglement distribution is on the same order of classical communication delay.


\appendices
\section{Proof of Theorem \ref{thm:rate_approximation}}\label{apd:proof_rate_approximation}

To shown the entanglement distribution rate is given by \eqref{eq:R_small_q}, we only need to determine the stationary distribution of the whole repeater chain system. The state of a repeater chain system with $2^n$ channels is represented recursively as follows. For $n=1$, the state is a scalar $S_1$, which represents the number of stored level 0 \ac{epr} pairs in the system. Note that with the entanglement swapping protocol, the stored \ac{epr} pairs are either $S_1$ pairs of $\ket{\Psi_{01}}$ or $S_1$ pairs of $\ket{\Psi_{12}}$. For $n>1$, the repeater chain can be divided into two sub repeater chains, each has $2^{n-1}$ channels, and let $S^{\mathrm{L}}_{n-1}$ and $S^{\mathrm{R}}_{n-1}$ denote the states of the left and right sub repeater chains, respectively. The state of the repeater chain is 
\begin{align*}
    S_n = (S^{\mathrm{L}}_{n-1}, S^{\mathrm{R}}_{n-1}, E_n^{\mathrm{L}}, E_n^{\mathrm{R}})
\end{align*}
where $E_n^{\mathrm{L}}$ and $E_n^{\mathrm{R}}$ denote the number of \ac{epr} pairs $\ket{\Psi_{0 2^{n-1}}}$ and $\ket{\Psi_{2^{n-1} 2^{n}}}$. Note that $E_n^{\mathrm{L}} \cdot E_n^{\mathrm{R}} = 0$. Let $\pi_{S_n}(s_n)$ denote the stationary distribution for the repeater that has $2^n$ channels, where state $s_n$ is one of the possible states. We next approximate $\pi_{S_n}$ for small $q$ recursively.
\begin{lemma}\label{lemma:stat_const_allocation}
The stationary distribution for the repeater state can be  approximated by $\underline{\pi}_{S_n}$ for small $q$, i.e.,
\begin{align*}
     \lim_{q\rightarrow 0}\pi_{S_n}(s_n)=\underline{\pi}_{S_n}(s_n)
\end{align*}
where $\underline{\pi}_{S_n}$ is given as follows. For $n=1$, $\underline{\pi}_{S_1}(s_1) = 1/(2B_0+1)$ for $s_1 = 1,2,\dotsc, B_0$ and $\underline{\pi}_{S_1}(s_1) = 2/(2B_0+1)$ for $s_1=0$. For $n>1$,
\begin{align*}
    \underline{\pi}_{S_n}(s_n) = \frac{\underline{\pi}_{S_{n-1}}(s^{\mathrm{L}}_{n-1})\underline{\pi}_{S_{n-1}}(s^{\mathrm{R}}_{n-1})}{2B_{n-1}+1}
\end{align*}
where $s^{\mathrm{L}}_{n-1}$ and $s^{\mathrm{R}}_{n-1}$ correspond to the states of the left right sub repeater chains.
\end{lemma}
\begin{IEEEproof}[Sketch of the Proof]
We prove this lemma by induction. For $n = 1$, one can easily verify that ${\pi}_{S_1}(s_1) = 2/(2B_0+1)$ for $s_1 = 1,2,\dotsc, B_0$ and ${\pi}_{S_1}(s_1) = 2/(2B_0+1)$ for $s_1=0$.

\begin{figure}[t]
\center	
\includegraphics[width=1\linewidth, draft=false]{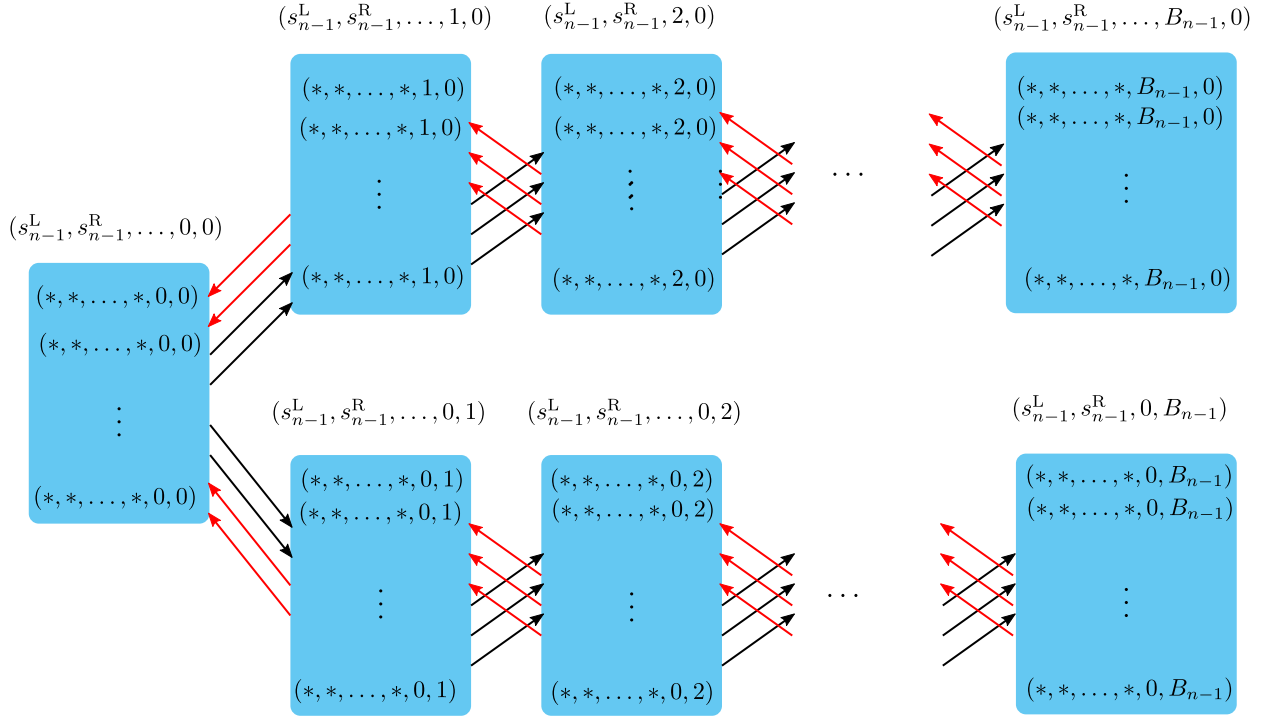}
\caption{Illustration of state transition for $S_n$. The symbol $*$ represents an arbitrary componentli of the state. }
\label{fig:state_transition}
\end{figure}

For $n>1$, the possible states can be grouped based on the last two components as shown in Fig.~\ref{fig:state_transition}. For sufficiently small $q$, the transitions between states from different groups can be ignored. Therefore, for $e_{n}^{\mathrm{L}}, e_{n}^{\mathrm{R}}\in\{0,1,2,\dotsc, B_{n-1}\}$ and $e_{n}^{\mathrm{L}}\cdot e_{n}^{\mathrm{R}}=0$,
\begin{align*}
    \lim_{q\rightarrow 0} \frac{\pi_{S_n}\big((s_{n-1}^{\mathrm{L}},s_{n-1}^{\mathrm{R}},e_{n}^{\mathrm{L}},e_{n}^{\mathrm{R}})\big)}{
    \pi_{S_n}\big((\tilde{s}_{n-1}^{\mathrm{L}},\tilde{s}_{n-1}^{\mathrm{R}},e_{n}^{\mathrm{L}},e_{n}^{\mathrm{R}})\big)
    }= \lim_{q\rightarrow 0}\frac{\pi_{S_{n-1}}(s_{n-1}^{\mathrm{L}})\pi_{S_{n-1}}(s_{n-1}^{\mathrm{R}})}{
    \pi_{S_{n-1}}(\tilde{s}_{n-1}^{\mathrm{L}})\pi_{S_{n-1}}(\tilde{s}_{n-1}^{\mathrm{R}})
    }.
\end{align*}
The induction hypothesis gives 
\begin{align*}
    \lim_{q\rightarrow 0}\frac{\pi_{S_{n-1}}(s_{n-1}^{\mathrm{L}})\pi_{S_{n-1}}(s_{n-1}^{\mathrm{R}})}{
    \pi_{S_{n-1}}(\tilde{s}_{n-1}^{\mathrm{L}})\pi_{S_{n-1}}(\tilde{s}_{n-1}^{\mathrm{R}})
    } =  \frac{\underline{\pi}_{S_{n-1}}(s_{n-1}^{\mathrm{L}})\underline{\pi}_{S_{n-1}}(s_{n-1}^{\mathrm{R}})}{
    \underline{\pi}_{S_{n-1}}(\tilde{s}_{n-1}^{\mathrm{L}})\underline{\pi}_{S_{n-1}}(\tilde{s}_{n-1}^{\mathrm{R}})
    }.
\end{align*}
Hence, 
\begin{align}\label{eq:proportion}
   \lim_{q\rightarrow 0} \frac{\pi_{S_n}\big((s_{n-1}^{\mathrm{L}},s_{n-1}^{\mathrm{R}},e_{n}^{\mathrm{L}},e_{n}^{\mathrm{R}})\big)}{
    \pi_{S_n}\big((\tilde{s}_{n-1}^{\mathrm{L}},\tilde{s}_{n-1}^{\mathrm{R}},e_{n}^{\mathrm{L}},e_{n}^{\mathrm{R}})\big)
    } =  \frac{\underline{\pi}_{S_{n-1}}(s_{n-1}^{\mathrm{L}})\underline{\pi}_{S_{n-1}}(s_{n-1}^{\mathrm{R}})}{
    \underline{\pi}_{S_{n-1}}(\tilde{s}_{n-1}^{\mathrm{L}})\underline{\pi}_{S_{n-1}}(\tilde{s}_{n-1}^{\mathrm{R}})
    }.
\end{align}

We then calculate the total probability for each group. By comparing the transitions between $(s_{n-1}^{\mathrm{L}},s_{n-1}^{\mathrm{R}},0,B_{n-1})$ and $(s_{n-1}^{\mathrm{L}},s_{n-1}^{\mathrm{R}},0,B_{n-1}-1)$,  we have
\begin{align*}
   \lim_{q\rightarrow 0} \sum_{s_{n-1}^{\mathrm{L}},s_{n-1}^{\mathrm{R}}} \pi_{S_n}\big((s_{n-1}^{\mathrm{L}},s_{n-1}^{\mathrm{R}},0,B_{n-1})\big) =  \lim_{q\rightarrow 0} \sum_{s_{n-1}^{\mathrm{L}},s_{n-1}^{\mathrm{R}}} \pi_{S_n}\big((s_{n-1}^{\mathrm{L}},s_{n-1}^{\mathrm{R}},0,B_{n-1}-1)\big).
\end{align*}
Similarly, we have
\begin{align*}
   &\lim_{q\rightarrow 0} \sum_{s_{n-1}^{\mathrm{L}},s_{n-1}^{\mathrm{R}}} \pi_{S_n}\big((s_{n-1}^{\mathrm{L}},s_{n-1}^{\mathrm{R}},0,B_{n-1})\big)\\
   & =  \lim_{q\rightarrow 0} \sum_{s_{n-1}^{\mathrm{L}},s_{n-1}^{\mathrm{R}}} \pi_{S_n}\big((s_{n-1}^{\mathrm{L}},s_{n-1}^{\mathrm{R}},0,B_{n-1}-1)\big)  =  \lim_{q\rightarrow 0} \sum_{s_{n-1}^{\mathrm{L}},s_{n-1}^{\mathrm{R}}} \pi_{S_n}\big((s_{n-1}^{\mathrm{L}},s_{n-1}^{\mathrm{R}},0,B_{n-1}-2)\big) \\
   &=\dotsc\\
   & =  \lim_{q\rightarrow 0} \sum_{s_{n-1}^{\mathrm{L}},s_{n-1}^{\mathrm{R}}} \pi_{S_n}\big((s_{n-1}^{\mathrm{L}},s_{n-1}^{\mathrm{R}},0,0)\big) =  \lim_{q\rightarrow 0} \sum_{s_{n-1}^{\mathrm{L}},s_{n-1}^{\mathrm{R}}} \pi_{S_n}\big((s_{n-1}^{\mathrm{L}},s_{n-1}^{\mathrm{R}},1,0)\big) \\
   & =  \lim_{q\rightarrow 0} \sum_{s_{n-1}^{\mathrm{L}},s_{n-1}^{\mathrm{R}}} \pi_{S_n}\big((s_{n-1}^{\mathrm{L}},s_{n-1}^{\mathrm{R}},2,0)\big) \\
   &=\dotsc\\
   & =  \lim_{q\rightarrow 0} \sum_{s_{n-1}^{\mathrm{L}},s_{n-1}^{\mathrm{R}}} \pi_{S_n}\big((s_{n-1}^{\mathrm{L}},s_{n-1}^{\mathrm{R}},B_n,0)\big).
\end{align*}
That means the total probability for each group is the same when $q$ goes to 0. Combined this with \eqref{eq:proportion}, we have the desired result.
\end{IEEEproof}

For $S_n$, let $\Set{S}_n$ denote the set of states that have nonzero transition rates to create an \ac{epr} $\ket{\Psi_{02^n}}$. Then
\begin{align}\label{eq:R_n_pq}
    R_{n}(p,q) = pq^{n} \sum_{s_n\in\Set{S}_n} \pi_{S_n}(s_n).
\end{align}
It is straightforward to see that $\sum_{s_1\in\Set{S}_1} \pi_{S_1}(s_1)=2B_0/(2B_0+1)$. We can calculate $\sum_{s_n\in\Set{S}_n} \pi_{S_n}(s_n)$:
\begin{align*}
    \sum_{s_n\in\Set{S}_n} \pi_{S_n}(s_n) &= \sum_{e^{\mathrm{R}}_{n}=1}^{B_{n-1}}\sum_{s^{\mathrm{L}}_{n-1}\in\Set{S}_{n-1},s^{\mathrm{R}}_{n-1}} \pi_{S_{n}}\big((s^{\mathrm{L}}_{n-1}, s^{\mathrm{R}}_{n-1},0, e^{\mathrm{R}}_{n})\big) \\
    &\quad +\sum_{e^{\mathrm{L}}_{n}=1}^{B_{n-1}}\sum_{s^{\mathrm{R}}_{n-1}\in\Set{S}_{n-1},s^{\mathrm{L}}_{n-1}} \pi_{S_{n}}\big((s^{\mathrm{L}}_{n-1}, s^{\mathrm{R}}_{n-1},e^{\mathrm{L}}_{n},0)\big).
\end{align*}
When $q$ goes to 0, we have
\begin{align*}
    \lim_{q\rightarrow 0}\sum_{s_n\in\Set{S}_n} 
    \pi_{S_n}(s_n)& = \sum_{s_n\in\Set{S}_n} 
    \underline{\pi}_{S_n}(s_n) 
    \\
    &=  \sum_{e^{\mathrm{R}}_{n}=1}^{B_{n-1}}
     \sum_{s^{\mathrm{L}}_{n-1}\in\Set{S}_{n-1}}
     \frac{\underline{\pi}_{S_{n-1}}(s^{\mathrm{L}}_{n-1})}{2B_{n-1}+1} +\sum_{e^{\mathrm{L}}_{n}=1}^{B_{n-1}}
     \sum_{s^{\mathrm{R}}_{n-1}\in\Set{S}_{n-1}}
     \frac{\underline{\pi}_{S_{n-1}}(s^{\mathrm{R}}_{n-1})}{2B_{n-1}+1} \\
     & =\frac{2B_{n-1}}{2B_{n-1}+1} \lim_{q\rightarrow 0} \sum_{s_{n-1}\in\Set{S}_{n-1}} 
    \pi_{S_{n-1}}(s_{n-1}).
\end{align*}
This gives a recursion, and
\begin{align*}
    \lim_{q\rightarrow 0}\sum_{s_n\in\Set{S}_n} 
    \pi_{S_n}(s_n)& =\frac{2B_{n-1}}{2B_{n-1}+1} \lim_{q\rightarrow 0} \sum_{s_{n-1}\in\Set{S}_{n-1}} 
    \pi_{S_{n-1}}(s_{n-1}) =\dotsc=\prod_{i=1}^{n-1}\frac{2B_i}{2B_i+1}\sum_{s_1\in\Set{S}_1} \pi_{S_1}(s_1) \\
    & = \prod_{i=0}^{n-1}\frac{2B_i}{2B_i+1}.
\end{align*}
Together with \eqref{eq:R_n_pq}, we have the desired result.

\section{Proof of Theorem \ref{thm:rate_approximation_cons}}\label{apd:proof_rate_approximation_cons}

We consider an auxiliary method that achieves a rate no greater than the cognitive memory allocation method, and show that this auxiliary method achieves a rate of $\underline{E}(p,q,B,i)$. In this auxiliary method, the doubling entanglement swapping protocol is used and higher level entanglement has priority to use memories. The following part is its key difference from the cognitive method: whenever a level $i$ \ac{epr} pair $\ket{\Psi_{(n-2^i)n}}$ is generated, two virtual qubits are stored from nodes $n-2^i+$ to node $n-1$ (one virtual qubit per link). These virtual qubits have higher priority to occupy the memory than entanglement with levels lower than $i$. Whenever a level $i$ \ac{epr} pair $\ket{\Psi_{(n-2^i)n}}$ is consumed, these virtual qubits are removed from the memory. This method achieves no greater rate the cognitive memory allocation method because the latter does not have virtual qubits occupying the memory.

The state of a repeater chain can be represented by $S_n$, the same way as in the proof of Theorem \ref{thm:rate_approximation}.  Let $\pi^b_{S_n}(s_n)$ denote the stationary distribution for the repeater that has $2^n$ channels, where state $s_n$ is one of the possible states and $b$ represents the number of memories per link. We next approximate $\pi^b_{S_n}(s_n)$ for small $q$ recursively.

\begin{lemma}
The stationary distribution for the repeater state can be  approximated by $\underline{\pi}_{S_n}$ for small $q$, i.e.,
\begin{align*}
     \lim_{q\rightarrow 0}\pi^b_{S_n}(s_n)=\underline{\pi}^b_{S_n}(s_n)
\end{align*}
where $\underline{\pi}^b_{S_n}$ is given as follows. For $n=1$, $\underline{\pi}^b_{S_1}(s_1) = 1/(2B+1)$ for $s_1 = 1,2,\dotsc, B$ and $\underline{\pi}^b_{S_1}(s_1) = 2/(2B+1)$ for $s_1=0$. For $n>1$,
\begin{align*}
    \underline{\pi}^b_{S_n}(s_n) = \underline{\pi}^{b-e^{\mathrm{L}}_{n}}_{S_{n-1}}(s^{\mathrm{L}}_{n-1})\underline{\pi}^{b-e_n^{\mathrm{R}}}_{S_{n-1}}(s^{\mathrm{R}}_{n-1})\frac{\prod_{j_0=0}^{\max\{e_n^{\mathrm{L}},e_n^{\mathrm{R}}\} f_{b-j_0, i-1}}}{(f_{b,i-1})^{\max\{e_n^{\mathrm{L}},e_n^{\mathrm{R}}\}+1}}\bigg[1+2\sum_{j =0}^{b-1}\frac{\prod_{j_0=0}^j f_{b-j_0, i-1}}{(f_{b,i-1})^{j+1}}\bigg]^{-1}
\end{align*}
where $s^{\mathrm{L}}_{n-1}$ and $s^{\mathrm{R}}_{n-1}$ correspond to the states of the left right sub repeater chains;  $e^{\mathrm{L}}_{n}$ and $e^{\mathrm{R}}_{n}$ correspond to the last two components of $s_n$.
\end{lemma}
The proof is similar to the proof of Lemma \ref{lemma:stat_const_allocation} and is omitted here.

For $S_n$, let $\Set{S}^b_n$ denote the set of states that have nonzero transition rates to create an \ac{epr} $\ket{\Psi_{02^n}}$. Then
\begin{align}\label{eq:E_b_pq}
    E(p,q,b,i) = pq^{i} \sum_{s_n\in\Set{S}^b_n} \pi^b_{S_n}(s_n).
\end{align}
It is straightforward to see that $\sum_{s_1\in\Set{S}^b_1} \pi^b_{S_1}(s_1)=2b/(2b+1)$. We can calculate $\sum_{s_n\in\Set{S}^b_n} \pi^b_{S_n}(s_n)$:
\begin{align*}
    \sum_{s_n\in\Set{S}^b_n} \pi^b_{S_n}(s_n) &= \sum_{e^{\mathrm{R}}_{n}=1}^{b}\sum_{s^{\mathrm{L}}_{n-1}\in\Set{S}^b_n,s^{\mathrm{R}}_{n-1}} \pi^b_{S_{n}}\big((s^{\mathrm{L}}_{n-1}, s^{\mathrm{R}}_{n-1},0, e^{\mathrm{R}}_{n})\big) \\
    &\quad +\sum_{e^{\mathrm{L}}_{n}=1}^{b}\sum_{s^{\mathrm{R}}_{n-1}\in\Set{S}^b_n,s^{\mathrm{L}}_{n-1}} \pi^b_{S_{n}}\big((s^{\mathrm{L}}_{n-1}, s^{\mathrm{R}}_{n-1},e^{\mathrm{L}}_{n},0)\big).
\end{align*}
When $q$ goes to 0, we have
\begin{align*}
    \lim_{q\rightarrow 0}\sum_{s_n\in\Set{S}^b_n} 
    \pi^b_{S_n}(s_n)& = \sum_{s_n\in\Set{S}^b_n} 
    \underline{\pi}^b_{S_n}(s_n) 
  \\
     & =\frac{f_{B,i}}{f_{B, i-1}} \lim_{q\rightarrow 0} \sum_{s_{n-1}\in\Set{S}^b_{n-1}} 
    \pi^b_{S_{n-1}}(s_{n-1}).
\end{align*}
This gives a recursion, and
\begin{align*}
    \lim_{q\rightarrow 0}\sum_{s_n\in\Set{S}_n} 
    \pi_{S_n}(s_n) = f_{B,i}.
\end{align*}
Together with \eqref{eq:E_b_pq}, we have the desired result.


\begin{thebibliography}{10}
\providecommand{\url}[1]{#1}
\csname url@samestyle\endcsname
\providecommand{\newblock}{\relax}
\providecommand{\bibinfo}[2]{#2}
\providecommand{\BIBentrySTDinterwordspacing}{\spaceskip=0pt\relax}
\providecommand{\BIBentryALTinterwordstretchfactor}{4}
\providecommand{\BIBentryALTinterwordspacing}{\spaceskip=\fontdimen2\font plus
\BIBentryALTinterwordstretchfactor\fontdimen3\font minus
  \fontdimen4\font\relax}
\providecommand{\BIBforeignlanguage}[2]{{%
\expandafter\ifx\csname l@#1\endcsname\relax
\typeout{** WARNING: IEEEtran.bst: No hyphenation pattern has been}%
\typeout{** loaded for the language `#1'. Using the pattern for}%
\typeout{** the default language instead.}%
\else
\language=\csname l@#1\endcsname
\fi
#2}}
\providecommand{\BIBdecl}{\relax}
\BIBdecl

\bibitem{WehElkHan:18}
S.~Wehner, D.~Elkouss, and R.~Hanson, ``Quantum {I}nternet: {A} vision for the
  road ahead,'' \emph{Science}, vol. 362, no. 6412, pp. 1--9, Oct. 2018.

\bibitem{PirLauOttBan:17}
S.~Pirandola, R.~Laurenza, C.~Ottaviani, and L.~Banchi, ``Fundamental limits of
  repeaterless quantum communications,'' \emph{Nat. Commun.}, vol.~8, no.
  15043, 2017.

\bibitem{BriDurCirZol:98}
H.~J. Briegel, W.~D\"ur, J.~I. Cirac, and P.~Zoller, ``Quantum repeaters: The
  role of imperfect local operations in quantum communication,'' \emph{Phys.
  Rev. Lett.}, vol.~81, no.~26, pp. 5932--5935, Dec. 1998.

\bibitem{JiaTayKhaLuk:07}
L.~Jiang, J.~M. Taylor, N.~Khaneja, and M.~D. Lukin, ``Optimal approach to
  quantum communication using dynamic programming,'' \emph{Proc. Natl. Acad.
  Sci. USA}, vol. 104, no.~44, pp. 17\,291--17\,296, Oct. 2007.

\bibitem{JiaTayNemMunMetLuk:09}
L.~Jiang, J.~M. Taylor, K.~Nemoto, W.~J. Munro, R.~V. Meter, and M.~D. Lukin,
  ``Quantum repeater with encoding,'' \emph{Phys. Rev. A}, vol.~79, no.~3, p.
  032325, Mar. 2009.

\bibitem{ZwePirDunBriDur:18}
M.~Zwerger, A.~Pirker, V.~Dunjko, H.~J. Briegel, and W.~D\"ur, ``Long-range big
  quantum-data transmission,'' \emph{Phys. Rev. Lett.}, vol. 120, no.~3, p.
  030503, 2018.

\bibitem{DaiPenWin:J20b}
W.~Dai, T.~Peng, and M.~Z. Win, ``Optimal remote entanglement distribution,''
  \emph{{IEEE} J. Sel. Areas Commun.}, vol.~38, no.~3, pp. 540--556, Mar. 2020.

\bibitem{PanKroTowTasJiaBasEngGuh:19}
M.~Pant, H.~Krovi, D.~Towsley, L.~Tassiulas, L.~Jiang, P.~Basu, D.~Englund, and
  S.~Guha, ``Routing entanglement in the quantum {I}nternet,'' \emph{npj
  Quantum Inf.}, vol.~5, no.~1, pp. 1--9, Mar. 2019.

\bibitem{ShcSchLoo:19}
E.~Shchukin, F.~Schmidt, and P.~van Loock, ``Waiting time in quantum repeaters
  with probabilistic entanglement swapping,'' \emph{Phys. Rev. A}, vol. 100,
  no.~3, p. 032322, Sep. 2019.

\bibitem{ShcLoo:21}
\BIBentryALTinterwordspacing
E.~Shchukin and P.~van Loock, ``Optimal entanglement swapping in quantum
  repeaters,'' \emph{{arXiv}}, 2021. [Online]. Available:
  \url{https://arxiv.org/pdf/2109.00793.pdf}
\BIBentrySTDinterwordspacing

\bibitem{JiaTayLuk:07b}
L.~Jiang, J.~M. Taylor, and M.~D. Lukin, ``Fast and robust approach to
  long-distance quantum communication with atomic ensembles,'' \emph{Phys. Rev.
  A}, vol.~76, no.~1, p. 012301, Jul. 2007.

\bibitem{CooBraElk:21}
\BIBentryALTinterwordspacing
T.~Coopmans, S.~Brand, and D.~Elkouss, ``Improved analytical bounds on delivery
  times of long-distance entanglement,'' \emph{{arXiv}}, 2021. [Online].
  Available: \url{https://arxiv.org/abs/2103.11454}
\BIBentrySTDinterwordspacing

\bibitem{BraCooElk:20}
S.~Brand, T.~Coopmans, and D.~Elkouss, ``Efficient computation of the waiting
  time and fidelity in quantum repeater chains,'' \emph{{IEEE} J. Sel. Areas
  Commun.}, vol.~38, no.~3, pp. 619--639, Mar. 2020.

\bibitem{GooElkWeh:21}
K.~Goodenough, D.~Elkouss, and S.~Wehner, ``Optimizing repeater schemes for the
  quantum internet,'' \emph{Phys. Rev. A}, vol. 103, no.~3, p. 032610, Mar.
  2021.

\bibitem{Pre:18}
J.~Preskill, ``Quantum computing in the {NISQ} era and beyond,''
  \emph{Quantum}, vol.~2, p.~79, 2018.

\bibitem{DhaPatKroGuh:21}
\BIBentryALTinterwordspacing
P.~Dhara, A.~Patil, H.~Krovi, and S.~Guha, ``Sub-exponential rate versus
  distance with time multiplexed quantum repeaters,'' \emph{{arXiv}}, 2021.
  [Online]. Available: \url{https://arxiv.org/pdf/2105.01002}
\BIBentrySTDinterwordspacing

\end{thebibliography}
\end{document}